# A note on optical activity and extrinsic chirality

Oriol Arteaga

Dep. Física Aplicada i Òptica, Univ. Barcelona, Barcelona 08028, Spain
oarteaga@ub.edu

The phenomenon of optical activity was first explained by A. J. Fresnel in 1822 [1], by supposing that right- and left-circularly polarized light propagate with different velocities in the medium. The expression proposed by Fresnel for optical activity has the form

$$\emptyset = \pi(n_L - n_R)l/\lambda , \qquad (1)$$

where $n_L$ and $n_R$ are, respectively, the (complex) indices of refraction of the medium for left- and right- circularly polarized light. $\lambda$ is the vacuum wavelength of light and $l$ the optical pathlength of the medium.

Unfortunately, Eq. (1) has too many times led to the overly simplistic assumption that optical activity can be measured by illuminating alternatively a material with left- and right- handed circular polarized light and analyzing the differential response. This simple and intuitive approach is in general incorrect and, surprising as it can be, it is not in agreement with Eq. (1).

The propagation of light in a material is a continuous process, and cannot be presented as a "one-shot" interaction. As light propagates in the medium, the polarization of the wave progressively changes or even disappears due to effects such as linear retardation, diattenuation or depolarization (Fig. 1, bottom). As a result, a beam of light, initially circularly polarized, changes its polarization as it propagates and, globally, it is refracted or absorbed with an index that does not correspond with any of the indices in Eq. (1). However, in an isotropic chiral medium the circular polarization is an eigenmode of propagation, and the emerging wave, still circularly polarized, will have refracted and absorbed according to $n_L$ or $n_R$ at any point of the optical pathlength (Fig. 1, top). This is the single case where the optical activity measurement method outlined above works. In all other situations it is advisable to go for complete polarimetric solutions based on the determination of the Mueller matrix [2].

The term *extrinsic chirality* was coined relatively recently and it is mainly associated to achiral nanostructured materials that show a distinct response when illuminated with left- and right-circular polarization [3-6]. Due this different response, and succumbing to the flawed interpretation of optical activity, it seems it has been quite widely (and incorrectly) considered that these materials are optically active. In most occasions, their distinct response to the handedness of circular polarization is consequence of the combination of linear retardation (intrinsic of the material or induced by illumination at oblique incidence) and linear diattenuation, which for nanostructured materials is usually large. There is no doubt that the optical response of these materials is interesting, but it should not be confused with optical activity as it does not arise from a real magneto-electric contribution. Complete polarimetry can be used to parse the different effects [7].

For the sake of clarity it is worth to recall that optical activity is not forbidden in some achiral point groups and it has been detected experimentally [8], but this is a different case from the situation we have summarized.

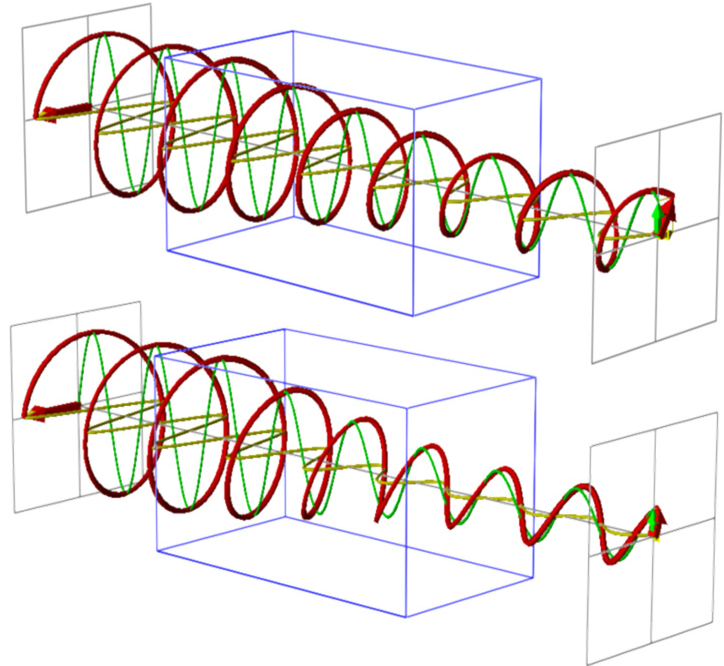

Fig 1. Propagation of circularly polarized light in an isotropic medium (top) and in anisotropic medium (bottom). The circular polarization is only preserved when there is isotropy.